\documentclass[%
reprint,
superscriptaddress,
preprintnumbers,
 amsmath,amssymb,
 aps,
prd,
]{revtex4-2}

\usepackage{graphicx}
\usepackage{dcolumn}
\usepackage{bm}
\usepackage[colorlinks = true,
            linkcolor = red,
            urlcolor  = blue,
            citecolor = red,
            anchorcolor = blue]{hyperref}
\usepackage[capitalise]{cleveref}
\usepackage[caption=false]{subfig}
\usepackage{multirow}
\usepackage{siunitx}

\begin{document}


\preprint{FERMILAB-PUB-25-0401-PPD}

\title{Sampling off-axis neutrino fluxes with the \\ short-baseline near detector
}

\newcommand{\ANL}{Argonne National Laboratory, Lemont, IL 60439, USA}
\newcommand{\Bern}{Universit\"{a}t Bern, Bern CH-3012, Switzerland}
\newcommand{\BNL}{Brookhaven National Laboratory, Upton, NY 11973, USA}
\newcommand{\UCSB}{University of California, Santa Barbara CA, 93106, USA}
\newcommand{\Campinas}{Universidade Estadual de Campinas, Campinas, SP 13083-970, Brazil}
\newcommand{\CTI}{Center for Information Technology Renato Archer, Campinas, SP 13069-901, Brazil}
\newcommand{\CERN}{CERN, European Organization for Nuclear Research 1211 Geneve 23, Switzerland, CERN}
\newcommand{\Chicago}{Enrico Fermi Institute, University of Chicago, Chicago, IL 60637, USA}
\newcommand{\CIEMAT}{CIEMAT, Centro de Investigaciones Energ\'{e}ticas, Medioambientales y Tecnol\'{o}gicas, Madrid E-28040, Spain}
\newcommand{\CSU}{Colorado State University, Fort Collins, CO 80523, USA}
\newcommand{\Columbia}{Columbia University, New York, NY 10027, USA}
\newcommand{\Edinburgh}{University of Edinburgh, Edinburgh EH9 3FD, United Kingdom}
\newcommand{\ABC}{Universidade Federal do ABC, Santo Andr\'{e}, SP 09210-580, Brazil}
\newcommand{\Alfenas}{Universidade Federal de Alfenas, Po\c{c}os de Caldas, MG 37715-400, Brazil}
\newcommand{\FNAL}{Fermi National Accelerator Laboratory, Batavia, IL 60510, USA}
\newcommand{\Florida}{University of Florida, Gainesville, FL 32611, USA}
\newcommand{\Granada}{Universidad de Granada, Granada E-18071, Spain}
\newcommand{\IIT}{Illinois Institute of Technology, Chicago, IL 60616, USA}
\newcommand{\Imperial}{Imperial College London, London SW7 2AZ, United Kingdom}
\newcommand{\SaoJose}{Instituto Tecnológico de Aeronáutica, São José dos Campos, SP 12228-900, Brazil}
\newcommand{\Kansas}{University of Kansas, Lawrence, KS 66045, USA}
\newcommand{\Lancaster}{Lancaster University, Lancaster LA1 4YW, United Kingdom}
\newcommand{\Liverpool}{University of Liverpool, Liverpool L69 7ZE, United Kingdom}
\newcommand{\LANL}{Los Alamos National Laboratory, Los Alamos, NM 87545, USA}
\newcommand{\LSU}{Louisiana State University, Baton Rouge, LA 70803, USA}
\newcommand{\Manchester}{University of Manchester, Manchester M13 9PL, United Kingdom}
\newcommand{\Michigan}{University of Michigan, Ann Arbor, MI 48109, USA}
\newcommand{\Minnesota}{University of Minnesota, Minneapolis, MN 55455, USA}
\newcommand{\Holyoke}{Mount Holyoke College, South Hadley, MA 01075, USA}
\newcommand{\Oxford}{University of Oxford, Oxford OX1 3RH, United Kingdom}
\newcommand{\Palermo}{Università degli Studi di Palermo, Dipartimento di Fisica e Chimica, I-90123 Palermo, Italy}
\newcommand{\Penn}{University of Pennsylvania, Philadelphia, PA 19104, USA}
\newcommand{\QueenMary}{Queen Mary University of London, London E1 4NS, United Kingdom}
\newcommand{\Rutgers}{Rutgers University, Piscataway, NJ, 08854, USA}
\newcommand{\Sheffield}{University of Sheffield, School of Mathematical and Physical Sciences, Sheffield S3 7RH, United Kingdom}
\newcommand{\Sussex}{University of Sussex, Brighton BN1 9RH, United Kingdom}
\newcommand{\Syracuse}{Syracuse University, Syracuse, NY 13244, USA}
\newcommand{\UCL}{University College London, London WC1E 6BT, United Kingdom}
\newcommand{\UTK}{University of Tennessee at Knoxville, TN 37996, USA}
\newcommand{\TAMU}{Texas A\&M University, College Station, TX 77843, USA}
\newcommand{\UTA}{University of Texas at Arlington, TX 76019, USA}
\newcommand{\Tufts}{Tufts University, Medford, MA, 02155, USA}
\newcommand{\VirginiaTech}{Center for Neutrino Physics, Virginia Tech, Blacksburg, VA 24060, USA}
\newcommand{\Warwick}{University of Warwick, Coventry CV4 7AL, United Kingdom}

\affiliation{\ANL}
\affiliation{\Bern}
\affiliation{\BNL}
\affiliation{\UCSB}
\affiliation{\Campinas}
\affiliation{\CTI}
\affiliation{\Chicago}
\affiliation{\CIEMAT}
\affiliation{\CSU}
\affiliation{\Columbia}
\affiliation{\Edinburgh}
\affiliation{\ABC}
\affiliation{\Alfenas}
\affiliation{\FNAL}
\affiliation{\Florida}
\affiliation{\Granada}
\affiliation{\IIT}
\affiliation{\Imperial}
\affiliation{\SaoJose}
\affiliation{\Kansas}
\affiliation{\Lancaster}
\affiliation{\Liverpool}
\affiliation{\LANL}
\affiliation{\LSU}
\affiliation{\Manchester}
\affiliation{\Michigan}
\affiliation{\Minnesota}
\affiliation{\Holyoke}
\affiliation{\Oxford}
\affiliation{\Palermo}
\affiliation{\Penn}
\affiliation{\QueenMary}
\affiliation{\Rutgers}
\affiliation{\Sheffield}
\affiliation{\Sussex}
\affiliation{\Syracuse}
\affiliation{\UTK}
\affiliation{\TAMU}
\affiliation{\UTA}
\affiliation{\Tufts}
\affiliation{\VirginiaTech}
\affiliation{\Warwick}
\affiliation{\UCL}

\author{P.~Abratenko} \affiliation{\Tufts}
\author{R.~Acciarri} \affiliation{\FNAL}
\author{C.~Adams} \affiliation{\ANL}
\author{L.~Aliaga-Soplin} \affiliation{\UTA}
\author{O.~Alterkait} \affiliation{\Tufts}
\author{R.~Alvarez-Garrote} \affiliation{\CIEMAT}
\author{D.~Andrade Aldana} \affiliation{\IIT}
\author{C.~Andreopoulos} \affiliation{\Liverpool}
\author{A.~Antonakis} \affiliation{\UCSB}
\author{L.~Arellano} \affiliation{\Manchester}
% ...

\collaboration{The SBND Collaboration}





\date{\today}

\begin{abstract}


    
    The short-baseline near detector (SBND), the near detector in the short-baseline neutrino program at Fermi National Accelerator Laboratory, is located just 110 m from the booster neutrino beam target.
    Thanks to this close proximity, relative to its 4~m $\times$ 4~m front face, neutrinos enter SBND over a range of angles from $0^{\circ}$ to approximately $1.6^{\circ}$, enabling the detector to sample variations in the neutrino flux as a function of angle---a technique known as precision reaction independent spectrum measurement (PRISM), referred to here as SBND-PRISM.
    In this paper, we show how muon- and electron-neutrino fluxes vary as a function of the neutrino beam axis angle and how this can be exploited to expand the physics potential of SBND.
    We make use of a model that predicts an angle-dependent electron-neutrino excess signal to illustrate this effect, such as $\nu_\mu \to \nu_e$ oscillations.
    We present how SBND-PRISM provides a method to add robustness against uncertainties in cross-section modeling and, more generally, uncertainties that do not depend on the spatial position of neutrino interaction inside the detector.
    The fluxes, along with their associated covariance matrices, are made publicly available with this publication. 
\end{abstract}

\maketitle

\tableofcontents

\section{Introduction}

The short-baseline near detector (SBND), a liquid argon neutrino detector located at Fermi National Accelerator Laboratory (Fermilab), receives accelerator neutrinos from Fermilab’s booster neutrino beam (BNB)~\cite{miniboone_flux}. SBND is the near detector of the short-baseline neutrino (SBN) experimental program~\cite{MicroBooNE:2015bmn}, alongside a far detector, ICARUS~\cite{icarus}. SBN was designed to conduct a highly sensitive multidetector search for sterile neutrinos at the eV mass scale. In addition, SBND will explore a broad physics program, from measurements of neutrino-argon interactions with a data sample of millions of events to searches for signatures of new physics beyond the Standard Model (BSM)~\cite{Machado:2019oxb}.

In this paper, we focus on a particular aspect of SBND: its proximity to the neutrino production source.
Thanks to this, neutrinos arrive at a range of off-axis angles with respect to the beam line axis that can be resolved by SBND.
In a two-body decay like $\pi^+ \rightarrow \mu^+ + \nu_\mu$, the neutrino energy in the laboratory frame depends on the angle between the pion momentum and the neutrino emission direction~\cite{mcdonald2001offaxisneutrinobeam}. On average, neutrinos emitted in the forward direction (small angles) carry more energy, while those emitted at larger angles (i.e., off axis) have lower energies.
This angular dependence leads to a correlation between off-axis position and neutrino energy, a key feature exploited by SBND-PRISM (precision reaction independent spectrum measurement), enabled by excellent position resolution. This feature allows probing different neutrino spectra within the same detector.
The neutrino-flux dependence on the off-axis angle is flavor and interaction dependent. For example, three-body decays, which contribute to the production of electron neutrinos, have a broader angular distribution, resulting in a smaller rate of decrease compared to muon neutrinos, which are almost exclusively produced in two-body decays.
We show how this can be leveraged to perform analyses that are robust against large uncertainties in cross-section modeling. The benefit here arises from disentangling uncertainties associated with the neutrino flux from those associated with neutrino-interaction cross sections.
We explore the use of this concept in improving the study of neutrino interactions and testing physics models that predict off-axis-angle-dependent signal distortions, such as $\nu_\mu \to \nu_e$ oscillations.
While $\nu_\mu \to \nu_e$ oscillations are not expected driven by standard oscillation mechanisms~\cite{Denton:2022een}, SBND is well situated to have sensitivity to new oscillations governed by a new oscillation length described by a mass splitting $\Delta m^2 \gtrsim 10$~eV$^2$~\cite{Machado:2019oxb}.

The technique of exploiting the off-axis angles of the neutrino in physics analyses is here called SBND-PRISM, in analogy to the similar concepts of nuPRISM~\cite{nuprism} and DUNE-PRISM~\cite{dunetdr}. However, while nuPRISM and DUNE-PRISM require physically moving the detector to sample different off-axis positions \footnote{Moving the detectors, or using dedicated off-axis detectors~\cite{MicroBooNE:2021gfj,NOvA:2024rov,Wood:2024jos}, can enable measurements at significantly larger angles than SBND can access.}, SBND-PRISM is enabled by the large solid angle as viewed by the protons striking the BNB target.
Using the angular dependence of the neutrino beam has been previously exploited, for example, in the T2K-INGRID detector~\cite{ingrid} leveraged the variation in neutrino energy spectra across its length to extract energy-dependent neutrino cross sections. Earlier, the CCFR experiment~\cite{ccfr} evaluated the neutrino energy in annular bins concentric with the beam axis.

In this paper, we first describe the BNB beam line and SBND in \cref{sec:detector}, then present the neutrino fluxes and discuss the SBND-PRISM technique in \cref{sec:fluxes}. In \cref{sec:applications} we describe some physics opportunities of SBND-PRISM and in \cref{sec:steriles} we illustrate how this technique can be applied to enhance sensitivity toward physics models that predict an off-axis-angle-dependent excess of electron-neutrino interactions, such as $\nu_\mu \to \nu_e$ oscillations.

\section{The BNB Beam line and SBND}
\label{sec:detector}

\begin{figure}[t]
    \includegraphics[width=0.48\textwidth]{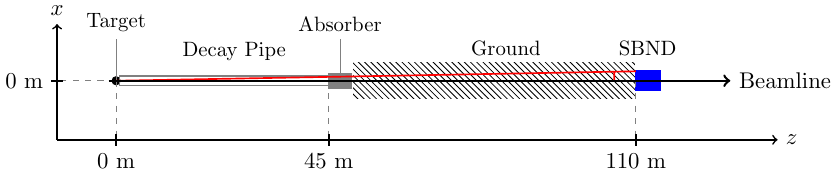}
    \caption{
        Illustration of the BNB beam line and SBND.
        The origin at \qty{0}{\meter} corresponds to the beginning of the target hit by the proton beam. The red line demonstrates the largest off-axis angle for neutrinos interacting in the upper edge of the SBND volume upstream face.}
    \label{fig:beamline}
\end{figure}

SBND is located along the BNB beam line at Fermilab, with the layout as shown in~\cref{fig:beamline}. Neutrinos at the BNB are created by colliding \qty{8}{\GeV} kinetic-energy protons on a beryllium target~\cite{miniboone_flux}. This collision produces secondary mesons, especially pions and kaons. The target is surrounded by an electromagnetic horn that generates a magnetic field to focus positive secondary particles and defocus the negative ones. The secondary particles then enter a 45-m region where most of the mesons decay, giving rise to the neutrino beam. The focusing of the positive secondaries gives rise to a neutrino-enhanced beam.
A 5-m-long absorber, positioned  at the end of the decay region, absorbs most of the surviving particles apart from the neutrinos.
The BNB runs parallel to the Earth's surface, $7$~m below ground, and the neutrinos produced in the BNB need to traverse an additional \qty{60}{\m} through the ground to reach SBND, which is located \qty{110}{\m} from the target.
In neutrino mode, the BNB consists primarily of $\nu_\mu$ ($\sim93.5$\%) with energy up to $\sim3$ GeV, followed by $\bar\nu_\mu$ ($\sim6$\%), $\nu_e$ ($\sim0.45$\%), and $\bar\nu_e$ ($\sim0.05$\%). 

\begin{figure}[t]
    \includegraphics[width=0.45\textwidth]{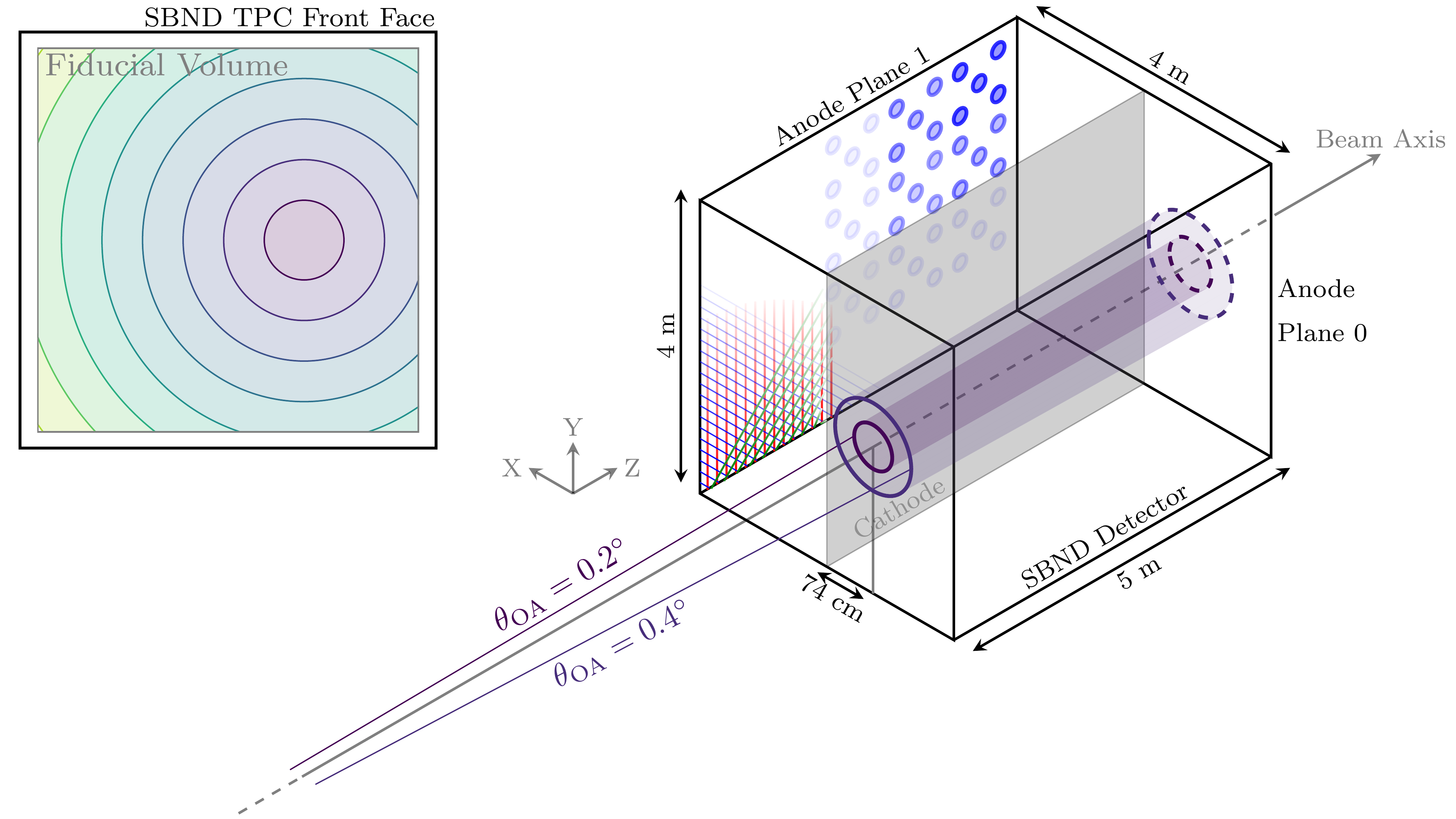}
    \caption{
        Illustration of the SBND position relative to the neutrino beam axis.
        The beam line crosses the detector \qty{74}{\cm} from the center. The gray panel shows the cathode, located at the center of the detector.
        As described in \cref{sec:fluxes}, the detector is divided into sections corresponding to different off-axis angles. Two of these regions are shown for $0.2^\circ$ and $0.4^\circ$. The radial thickness of the off-axis slices is approximately \qty{40}{\cm}.
        The upper-left inset shows all of the off-axis regions as they appear on the front face of the detector.
        Detector components on one anode plane (TPC wires in criss-crossed blue, red, and green lines, and photodetectors in blue circles) are shown for reference.}
    \label{fig:detector}
\end{figure}

The BNB beam line crosses SBND at \qty{74}{\cm} from the center vertical plane as shown in \cref{fig:detector}. The detector is a double-drift single-phase liquid argon time projection chamber (LArTPC)~\cite{lartpc}, with a central shared cathode and two anodes, one on each side of the detector.  Each TPC is \qty{5}{\m} long, \qty{4}{\m} high, and \qty{2}{\m} wide in the drift direction---for a total width of \qty{4}{\m}.

When a neutrino interacts within the SBND volume, it produces particles that traverse the liquid argon. Those which are charged create ionization trails and prompt ultraviolet scintillation light.
The ionization electrons drift horizontally, perpendicular to the beam direction, under an electric field of \qty{500}{V/cm}, moving toward the two anode planes.
Each anode is equipped with three planes of sense wires~\cite{Acciarri_2020} to detect this charge. The scintillation light is captured by an advanced photon detection system, which combines photomultiplier tubes and X-ARAPUCA devices
located behind the anode planes, complemented by highly reflective panels coated with a wavelength shifter on the cathode to maximize light collection~\cite{SBNDLight}.

SBND uses a right-handed coordinate system with the $x$ axis along the drift direction, the $y$ axis along the vertical, and the $z$ axis along the beam line direction, as shown in \cref{fig:detector} ($x=0$ corresponds to the cathode plane, $y = 0$ the middle of the TPC, and $z = 0$ the TPC front face). The intrinsic spatial resolution of the SBND TPC is given by the anode wire spacing in the $y$ and $z$ directions, and by the time-digitization frequency in the $x$ direction. With a wire spacing of \qty{3}{\mm} \cite{Acciarri_2020} and a sampling frequency of \qty{2}{MHz}, the detector intrinsic position resolution is less than \qty{3}{\mm} in $y$ and $z$, and \qty{0.35}{\mm} in $x$.

SBND started collecting neutrino data in December 2024, and is expected to accumulate an exposure of $10^{21}$ protons on target (POT) over three years of data taking, amounting to nearly 10 million neutrino-argon interaction events in the full SBND instrumented volume.

\section{Off-Axis Fluxes and the SBND-PRISM Technique}
\label{sec:fluxes}

\begin{figure*}
\subfloat[]{
  \includegraphics[width=0.66\columnwidth]{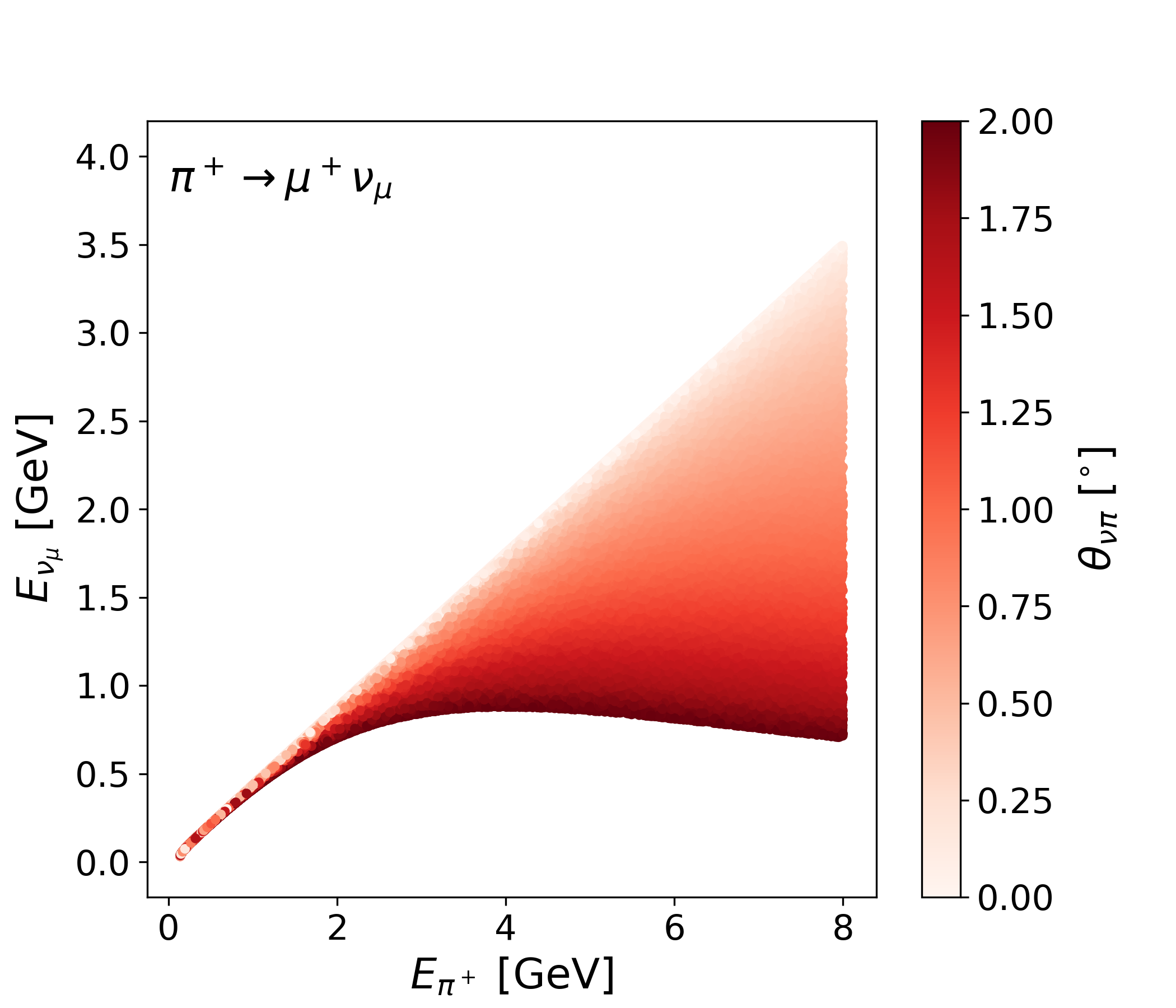}
  \label{fig:prism:pi}
}
\subfloat[]{
  \includegraphics[width=0.66\columnwidth]{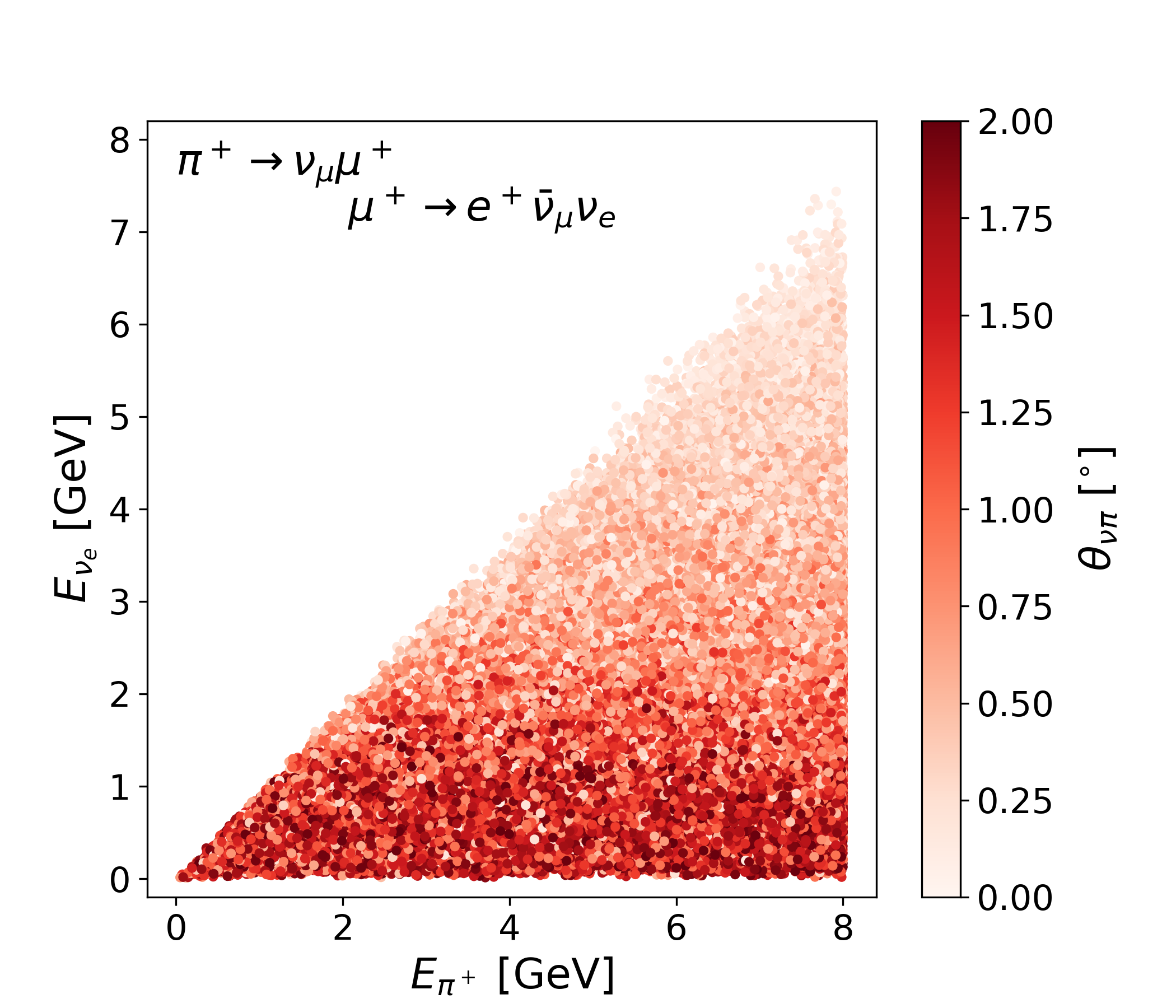}
  \label{fig:prism:mu}
}
\subfloat[]{
  \includegraphics[width=0.66\columnwidth]{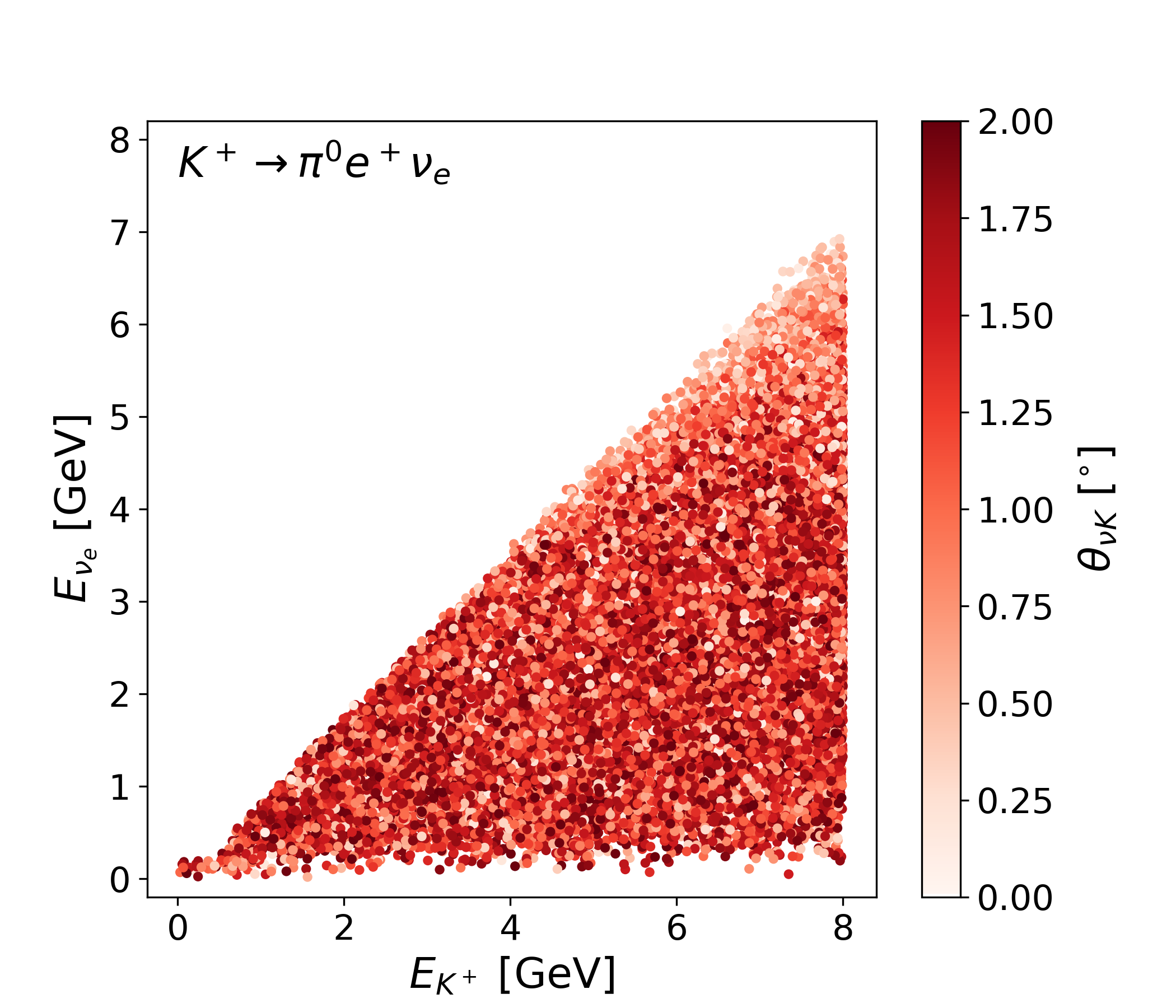}
   \label{fig:prism:k}
}
\caption{{\label{fig:prism}
\protect\subref{fig:prism:pi}
Energy carried by the neutrino in the two-body decay $\pi^+\to \mu^+\nu_\mu$ as a function of the pion energy for different decay angles. The $z$-axis scale is the angle between the incoming meson and the neutrino.
\protect\subref{fig:prism:mu} Energy carried by the electron neutrino in the three-body decay $\mu^+\to e^+\bar{\nu}_\mu\nu_e$, where the $\mu^+$ comes from a $\pi^+$ decay, as a function of the initial pion energy.
\protect\subref{fig:prism:k} Energy carried by the neutrino in the three-body decay $K^+\to \pi^0 e^+\nu_e$ as a function of the kaon energy.
\protect\subref{fig:prism:pi} shows a clear dependence on the angle. This dependence is also observed in \protect\subref{fig:prism:mu} although is less strong in this three-body decay. Finally, the dependence is largely lost in the kaon three-body decay shown in \protect\subref{fig:prism:k}.
}}
\end{figure*}

Upon collision of the 8~GeV protons on the BNB target, a large number of neutral and charged mesons is produced, which are mostly pions and kaons.
While neutral mesons decay quickly, the neutrino flux at SBND is dominated by the products of primary and secondary decays of the relatively long-lived charged pions and kaons. Muon and electron neutrinos come predominantly from two- and three-body decays, respectively. The main decays are:
\begin{align}
   \pi^+ & \to \mu^+ + \nu_\mu, \\
   K^+   & \to \mu^++\nu_\mu, \\
   K^+   & \to \pi^0 + e^+ + \nu_e, \\
   \mu^+ & \to e^+ + \nu_e + \bar{\nu}_\mu.
\end{align}
%


In \cref{fig:prism}, we examine the correlation between neutrino energy, neutrino angle, and the energy of the parent particles. Neutrino parents produced in the BNB are largely collinear with the beamline, with 90\% of them emitted at angles less than 3 degrees relative to the beam direction.
Due to the two-body kinematics of its production mode, the energy of the muon neutrino can be calculated knowing its angle and the parent energy. This can be observed in \cref{fig:prism:pi}, which is made by calculating analytically the neutrino kinematics from the pion decay.

The situation differs for electron neutrinos. Below $\sim 1$ GeV, $\nu_e$ are mainly produced in $\mu^+ \to e^+ \bar{\nu}_\mu \nu_e$ decays, while above this energy they primarily originate from $K^+ \to \pi^0 e^+ \nu_e$ \cite[Figs.~29, 30]{miniboone_flux}. In these three-body decays, the energy–angle correlation is governed by the mass differences between the particles in the initial and final states.
\cref{fig:prism:mu,fig:prism:k} show the cases for muon and kaon decay respectively, and have been made with Geant4~\cite{geant4} and the equations in \cite[Sec.~III.F]{miniboone_flux}, by generating muons and kaons uniformly between 0 and 8~GeV. For the muon decay, since most muons come from $\pi^+$ decays, \cref{fig:prism:mu} shows the correlation between the energy of the initial $\pi^+$ and the resulting electron neutrino from the decay chain $\pi^+ \to \nu_\mu + \mu^+$ followed by $\mu^+ \to e^+ \bar{\nu}_\mu \nu_e$. A  clear energy–angle correlation persists even in this more complex decay sequence.
The corresponding dependence for kaon decays is shown in \cref{fig:prism:k}. The larger mass difference between the kaon and its decay products
enables the resulting neutrino to carry more energy, thereby washing out the energy-angle correlation.
Compared to muon neutrinos, electron neutrinos exhibit some correlation between neutrino energy and parent particle energy at low energies---where they predominantly originate from muon decay---but this correlation becomes significantly weaker at higher energies, where kaon decays are the dominant source.

\begin{figure*}
    \subfloat[]{%
      \includegraphics[width=\columnwidth]{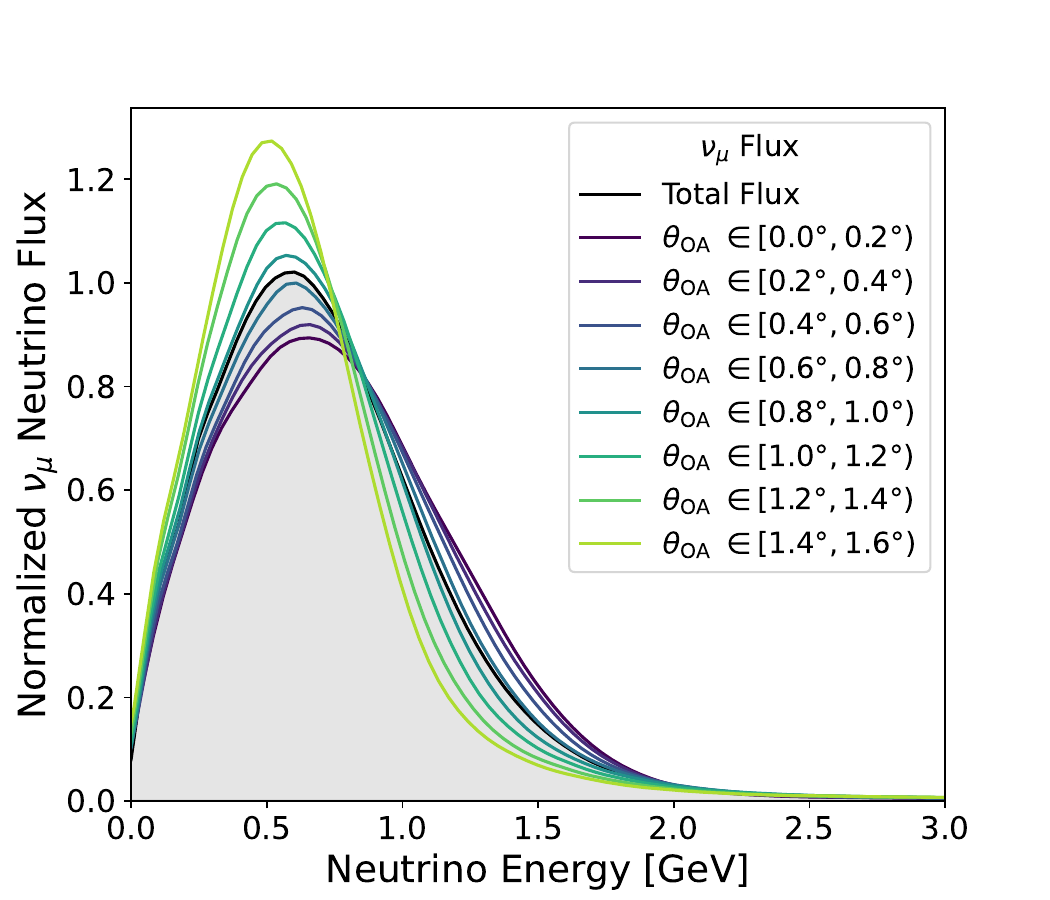}%
      \label{fig:prism_flux:numu}
    }
    \subfloat[]{%
      \includegraphics[width=\columnwidth]{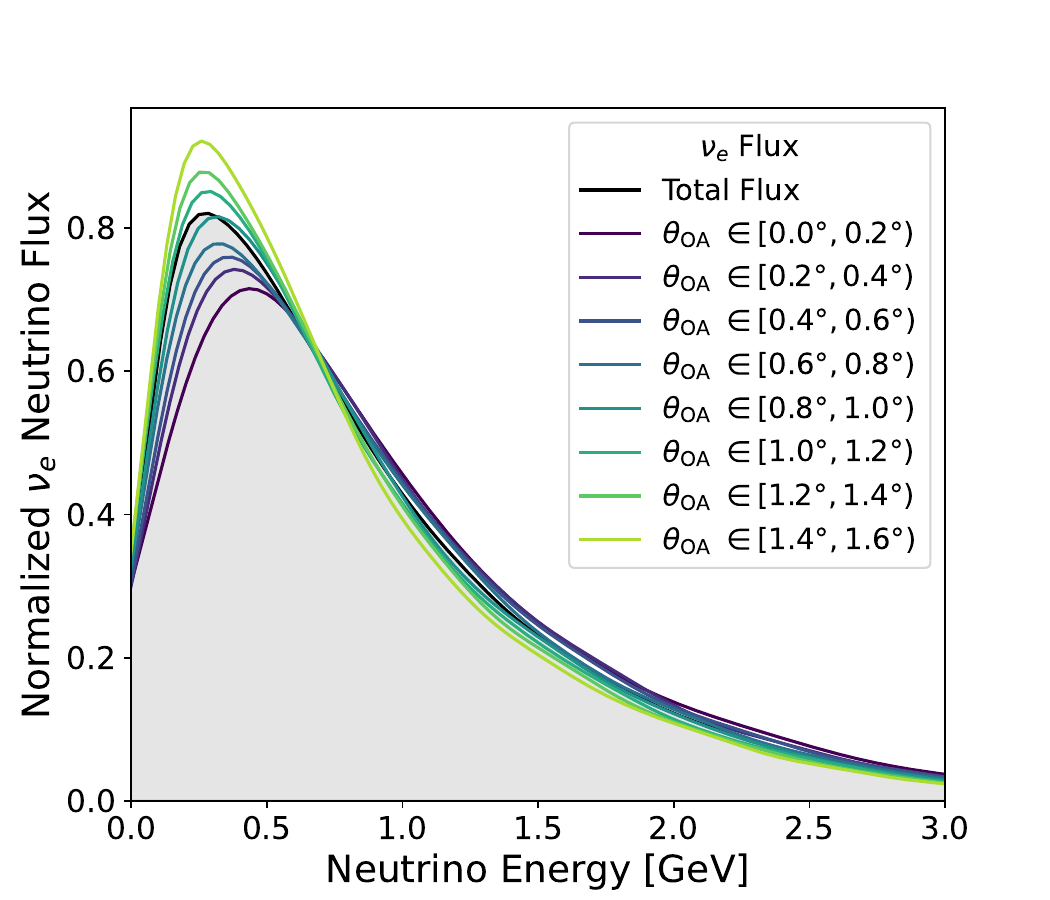}%
       \label{fig:prism_flux:nue}
    }
    \caption{{\label{fig:prism_flux}
        \protect\subref{fig:prism_flux:numu} Muon- and \protect\subref{fig:prism_flux:nue} electron-neutrino fluxes at SBND for different off-axis angles. Distributions are area normalized.}}
\end{figure*}

The previous considerations allow us to appreciate quantitatively the different dependence of the $\nu_\mu$ and $\nu_e$ energy spectra at SBND for different off-axis angles. To better study this effect,  we begin by dividing the SBND volume into several regions, each of which corresponds to specific values of the neutrino off-axis angle ($\theta_\text{OA}$). 

Since we cannot determine the exact location along the beam line where the neutrino was produced, we can only infer $\theta_\text{OA}$ based on where the neutrino interacts within the detector. 
We define $\theta_\text{OA}$ as the angle between two lines: the beam line axis and a line extending from the target to the neutrino interaction point within the detector. 
For instance, the red line in \cref{fig:beamline} illustrates $\theta_\text{OA}$ for a neutrino interacting in the upper edge of the SBND volume upstream face. 
With this definition of $\theta_\text{OA}$, we conceptually divide SBND into eight regions. 

As illustrated in \cref{fig:detector}, the first region is defined by a truncated cone centered on the beamline axis and contains neutrinos interacting inside this region that have $\theta_\text{OA}$ less than 0.2 degrees. 
The second region is defined to contain neutrinos that have $\theta_\text{OA}$ between 0.2 and 0.4 degrees. 
Six other regions are constructed in a similar manner, for a total coverage from 0 to 1.6 degrees off axis.
$\theta_\text{OA}$ is a practical definition and it differs from the decay angle $\theta_{\nu p}$ of the neutrino with respect to the parent particle $p = \pi, \mu, K$. The relationship between $\theta_\text{OA}$ and $\theta$ depends on where the neutrino is produced. The standard deviation of $\theta_\text{OA} - \theta_{\nu p}$ is around $1 ^\circ$.
For the studies in this paper, we only consider neutrino events in a fiducial volume defined as $-183.5 < x < 183.5$ cm, $-185.0 < y < 185.0$ cm, and $40.0 < z < 460.0$ cm, on the left in \cref{fig:detector}. The radial thickness of the off-axis slices is approximately \qty{40}{\cm}, significantly larger than the expected vertex resolution after calibration. For current reconstruction performance in similar LArTPCs, vertex resolution has been demonstrated to be better than \qty{1}{\cm} \cite{MicroBooNE:2020sar} after correcting for various detector effects, including the leading cause of smearing due to the space-charge effect~\cite{Adams_2018}.

The flux of neutrinos at SBND is simulated using the framework built by the MiniBooNE Collaboration~\cite{miniboone_flux} and adapted for SBND’s location along the beam line.
In \cref{fig:prism_flux}, we show the $\nu_\mu$ \protect\subref{fig:prism_flux:numu} and $\nu_e$ \protect\subref{fig:prism_flux:nue} area-normalized energy spectra for the different off-axis regions.
To aid visualization, the flux spectrum is shown as a smooth curve obtained using a kernel density estimation of the simulated neutrino energy distribution. The underlying binned flux histograms used in the analysis, together with their uncertainties, are provided as Supplementary Material~\cite{suppl}.
The impact of SBND-PRISM is evident in the observed fluxes: the $\nu_\mu$ energy spectrum changes noticeably with off-axis angle, with the mean energy shifting toward lower values and the distributions becoming narrower as the angle increases. This reflects its origin in two-body decays, which strongly correlate the neutrino energy with the parent meson direction, making it highly sensitive to the off-axis position.
In contrast, the effect on the $\nu_e$ flux is more subtle: it is most evident at low energies, where it is dominated by muon three-body decays, and less pronounced at higher energies, where kaon three-body decays dominate.


\begin{figure*}
    \subfloat[]{%
      \includegraphics[width=\columnwidth]{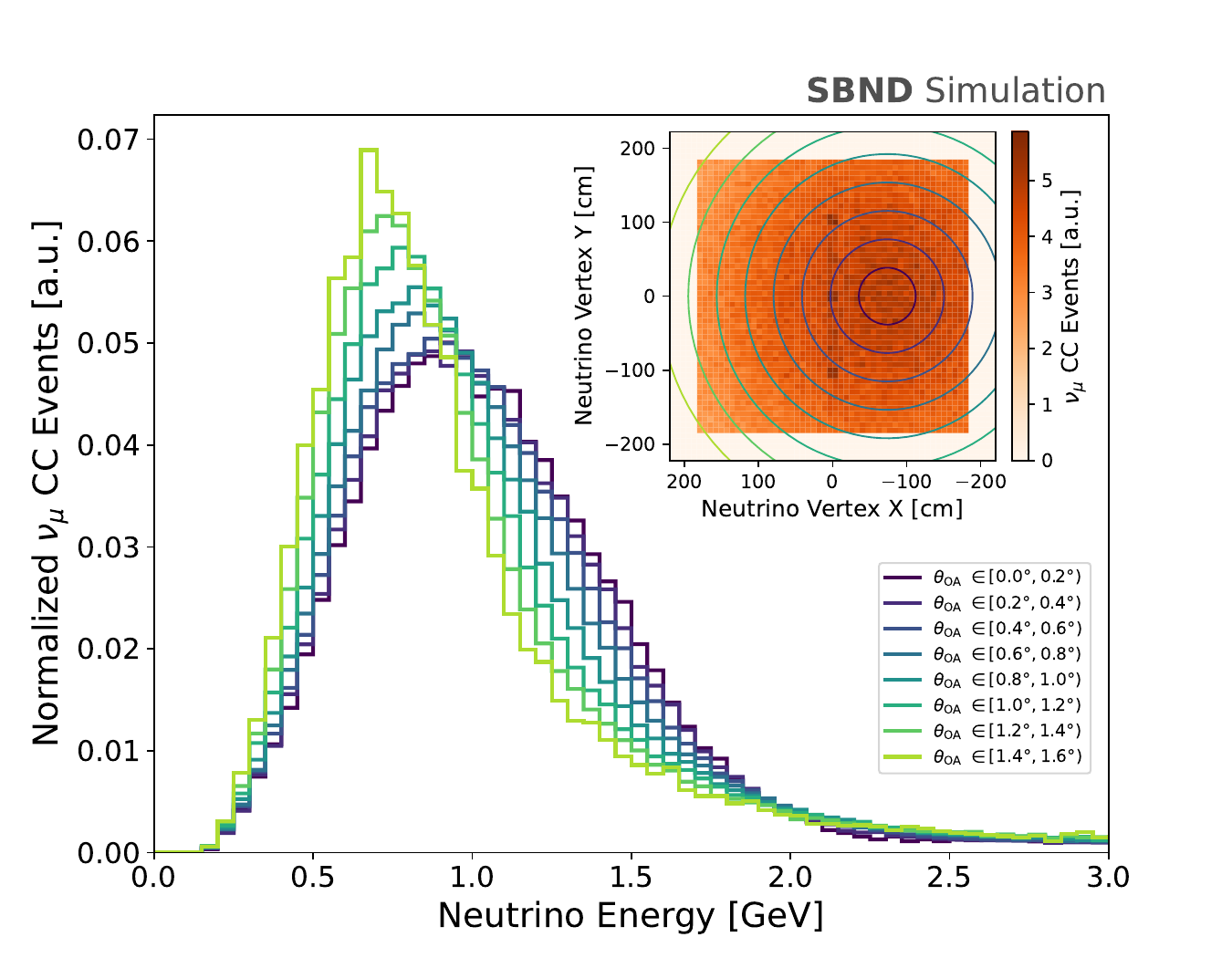}%
      \label{fig:prism_events:numu}
    }
    \subfloat[]{%
      \includegraphics[width=\columnwidth]{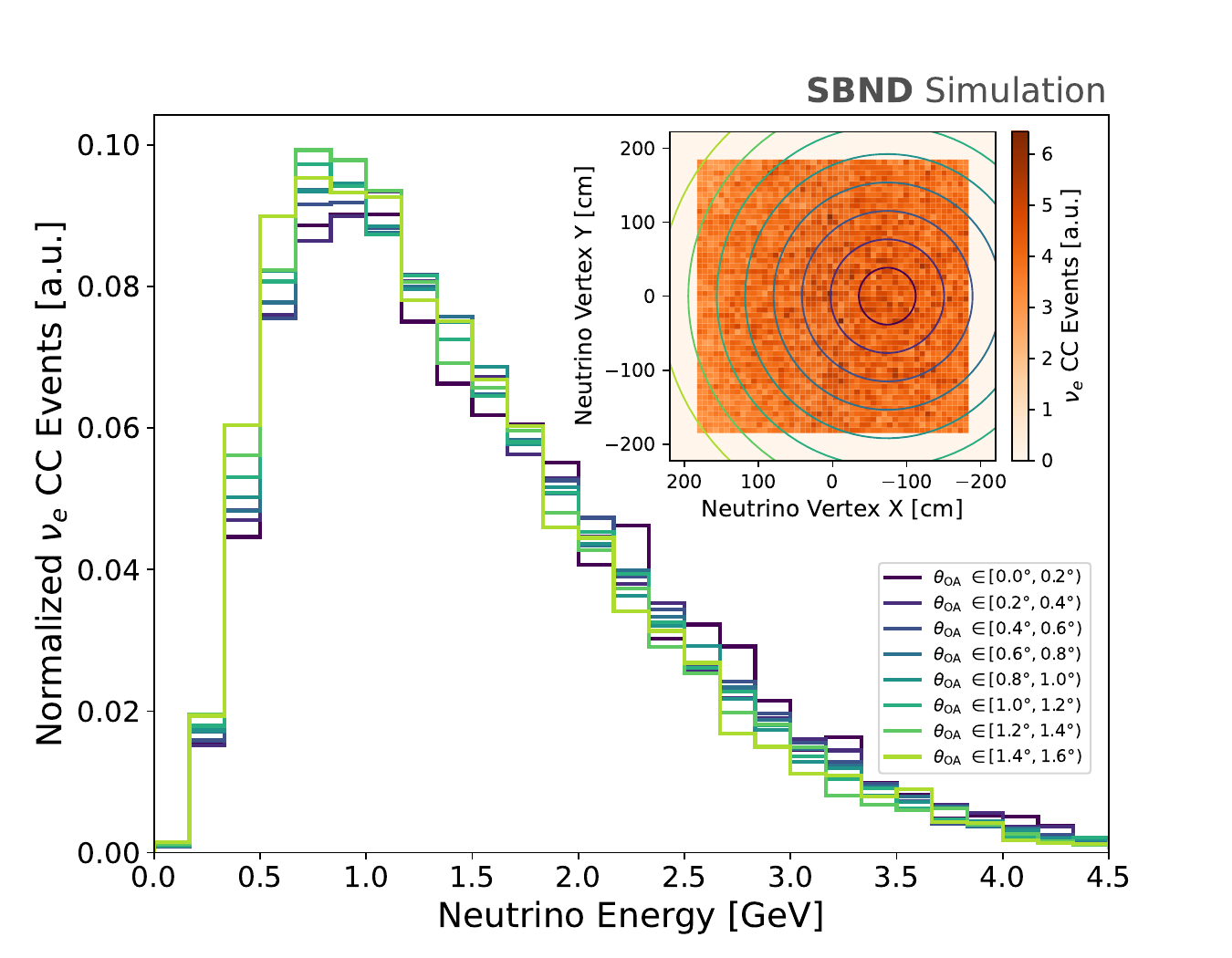}%
       \label{fig:prism_events:nue}
    }
    \caption{{\label{fig:prism_events}
        \protect\subref{fig:prism_events:numu} Muon- and \protect\subref{fig:prism_events:nue} electron-neutrino expected CC event distributions at SBND for different off-axis angles, from a simulation based on the GENIE \texttt{v3\_04\_02 AR23\_20i\_00\_000} neutrino generator. The insets show the location of the neutrino interaction vertices in the X-Y view. Distributions are normalized to unit area to highlight differences in spectral features.
        }}
\end{figure*}

It is important to note that SBND does not measure the neutrino flux, but a convolution of flux and cross section. 
As the cross section grows with energy, the spread at low energies for different fluxes gets washed out.
\cref{fig:prism_events} shows the $\nu_\mu$ \protect\subref{fig:prism_events:numu} and $\nu_e$ \protect\subref{fig:prism_events:nue} charged current (CC) neutrino events in the fiducial volume in each off-axis region, generated with the \texttt{GENIE} neutrino event generator~\cite{Andreopoulos:2009rq}, version \texttt{v3\_04\_02 AR23\_20i\_00\_000} and the full detector geometry description. 
The $\nu_\mu$ event rate per unit area decreases at larger off-axis angles, as shown in the inset of \cref{fig:prism_events:numu}, where the position of the vertices of the $\nu_\mu$ interactions projected on the front face of the detector is reported. This behavior is driven by the forward-focused kinematics of $\pi^+\to\mu^+\nu_\mu$ decays, which produce fewer and lower-energy neutrinos at larger off-axis angles.
Moving toward more off-axis regions, the observed $\nu_\mu$ energy spectrum narrows and peaks at a lower energy. 
There is a \qty{200}{\MeV} difference in the mean energy between the on-axis and the most off-axis region. 
The $\nu_e$ energy distributions also change, but they are less affected by the off-axis position (see \cref{fig:prism_events:nue}). 
The spatial distribution of $\nu_e$ events is 
more uniform (inset of \cref{fig:prism_events:nue}) compared to the $\nu_\mu$ events, which present a marked peak at the center of the beam line. 
As already discussed, this spatial difference is caused by the two- versus three-body decay of the mesons producing the neutrinos.

In \cref{tab:prism_event_table}, the expected numbers of $\nu_\mu$ CC, $\nu_e$ CC, and neutral-current (NC) events in the chosen fiducial volume for each off-axis region are shown. Large event statistics are expected for an exposure of $10^{21}$~POT in all off-axis regions.

\begin{table}[b]
\caption{
Expected number of $\nu_\mu$ CC, $\nu_e$ CC, and NC event rates in each off-axis region of SBND from a Monte Carlo simulation with the \texttt{GENIE} generator and for an exposure of $10^{21}$~POT. Events are also shown divided by the area of the detector front surface that subtends a particular off-axis region. The area that each off-axis region occupies on the front face of the detector is shown in the second column. }
\begin{ruledtabular}
\begin{tabular}{cccccccc}
\multirow{2}{*}{$\theta_\text{OA}$} & \multirow{2}{*}{Areas}  & \multicolumn{2}{c}{$\nu_\mu$ CC} & \multicolumn{2}{c}{$\nu_e$ CC} & \multicolumn{2}{c}{$\nu$ NC} \\
                                    &        & \multicolumn{2}{c}{Events}  & \multicolumn{2}{c}{Events}              & \multicolumn{2}{c}{Events}  \\
          deg                       & m$^2$ & $\times 10^3$ &  / cm$^2$        & $\times 10^3$ & / cm$^2$       & $\times 10^3$ & / cm$^2$ \\
\colrule
$[0.0, 0.2]$ & 0.46 & 192 & 41 & 1.25 & 0.27 & 74.7 & 16.1 \\
$[0.2, 0.4]$ & 1.39 & 557 & 40 & 3.71 & 0.27 & 222.2 & 16.0 \\
$[0.4, 0.6]$ & 2.29 & 869 & 38 & 5.89 & 0.26 & 349.2 & 15.2 \\
$[0.6, 0.8]$ & 2.61 & 925 & 36 & 6.48 & 0.25 & 375.7 & 14.4 \\
$[0.8, 1.0]$ & 2.89 & 918 & 32 & 6.69 & 0.24 & 373.7 & 12.9 \\
$[1.0, 1.2]$ & 1.89 & 524 & 28 & 4.08 & 0.21 & 213.3 & 11.3 \\
$[1.2, 1.4]$ & 1.49 & 352 & 24 & 2.89 & 0.19 & 145.3 & 9.7 \\
$[1.4, 1.6]$ & 5.21 & 92  & 18 & 0.79 & 0.15 & 38.7 & 7.4 \\
\colrule
Total & 13.58 & 4,428 & 33 & 31.78 & 0.24 & 1,792.8 & 3.2\\
\end{tabular}
\end{ruledtabular}
\label{tab:prism_event_table}
\end{table}


\section{Physics Opportunities of SBND-PRISM}
\label{sec:applications}

\begin{figure}
    \centering
        \includegraphics[width=0.45\textwidth]{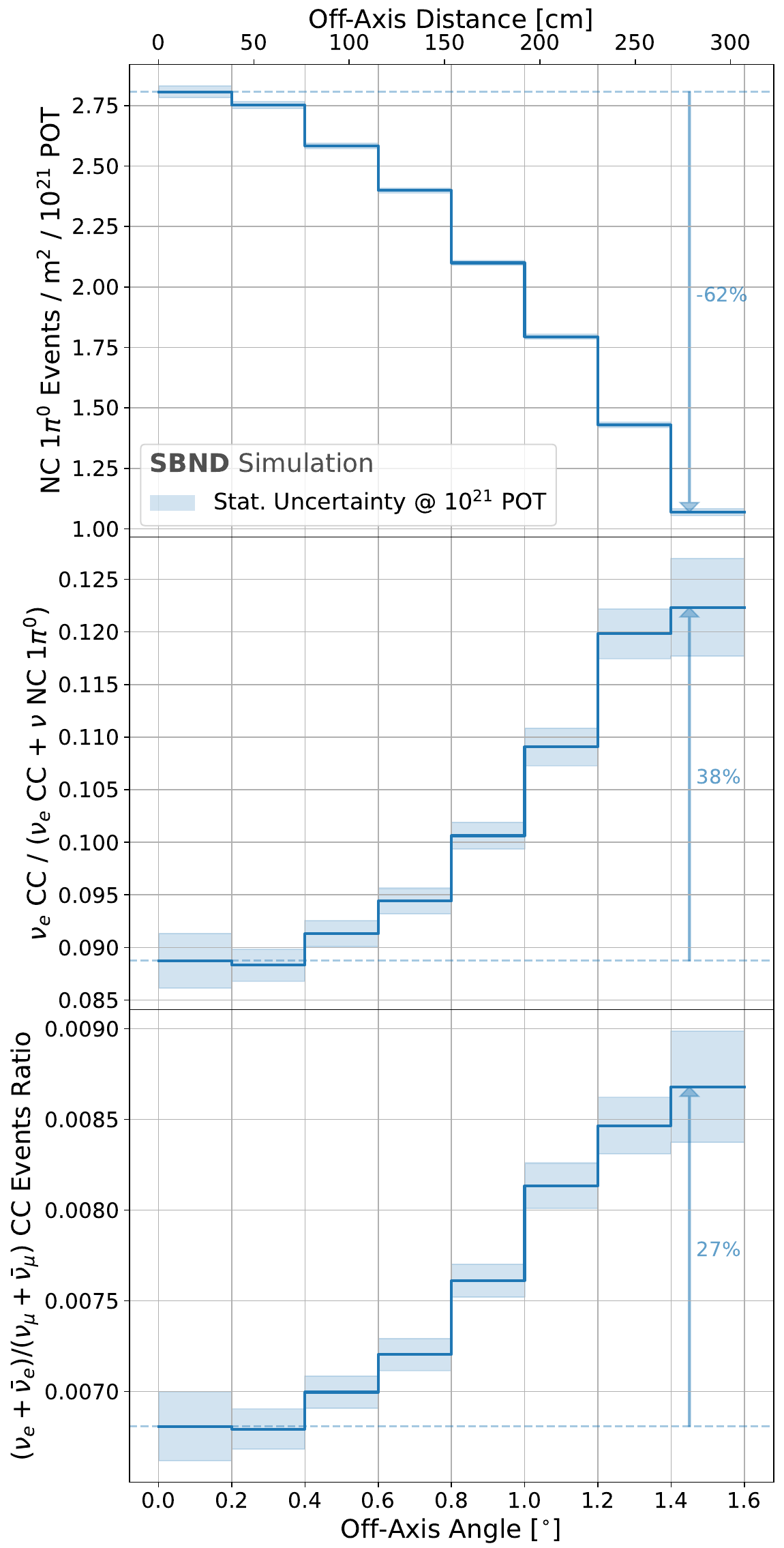}
        \caption{\label{fig:ratios}
        \emph{Top:} Number of NC neutrino events per $m^2$ with one $\pi^0$ in the final state in SBND as a function of the off-axis angle for an exposure of $10^{21}$ POT. 
        \emph{Middle:} Ratio between the number of CC electron-neutrino events and the sum of the number of electron-neutrino CC events and the number of NC events with one $\pi^0$ in the final state in SBND, as a function of the off-axis angle.
        \emph{Bottom:} Ratio between numbers of electron and muon neutrino CC events in SBND as a function of the off-axis angle.
        Each bin corresponds to an off-axis region; the light-blue bands indicate the statistical uncertainty corresponding to an exposure of $10^{21}$ POT.}
\end{figure}

SBND-PRISM offers several promising opportunities for enhancing the sensitivity of both BSM searches and Standard Model measurements in SBND, some of which are highlighted here. Others are currently being explored, with additional possibilities likely to emerge as more analysis tools and techniques  are developed.

The first area of opportunity that we have identified for SBND-PRISM is enabling further tests of neutrino-nucleus interactions.

By measuring neutrino interactions across different off-axis regions, we gain access to a range of neutrino energy spectra, enabling studies of cross-section energy dependence and nuclear effects. These variations allow us to test the consistency of data with different cross-section models and explore the relationship between neutrino energy and final-state kinematics by measuring differential cross sections in lepton and hadron variables. Additionally, combining fluxes from multiple regions can help isolate cleaner samples of specific interaction channels, such as quasielastic scattering, pion production, or interactions on correlated nucleons, providing a powerful handle on underlying nuclear effects.

As a second area of opportunity, SBND-PRISM is a valuable tool to study and mitigate backgrounds for certain analyses.
Electron-neutrino CC analyses and BSM searches looking for photons or electron-positron pairs in the final state~\cite{Bertuzzo:2018itn, Bertuzzo:2018ftf,Ballett:2018ynz,deGouvea:2018cfv, Arguelles:2018mtc,Ballett:2019pyw, Abdallah:2020vgg,Dutta:2021cip, ub1, ub2, ub3, ub4, t2k, pd} are affected by a large background from NC neutrino interactions producing one $\pi^0$ in the final state and no hadronic activity. 
This is because NC~$\pi^0$ events produce electromagnetic showers in the detector that can mimic the signature of CC $\nu_e$ events if one of the two photons from the $\pi^0$ decay is below detection threshold or escapes the detector. 

Figure~\ref{fig:ratios} (top) shows that the number of NC neutrino events with one $\pi^0$ in the final state decreases rapidly moving from on axis to off axis (a reduction of up to 62\%). The numbers in the plot have been divided by the annulus areas (see left in \cref{fig:detector}) to compare events in different off-axis regions.
This  behavior can be used to assess the effectiveness of the NC $\pi^0$ tagging algorithms employed in an analysis and can be exploited as a background mitigation strategy in several BSM searches. 
As a strategy to characterize or mitigate background in the analysis of $\nu_e$ events, one can use the fact that in SBND, the numbers of $\nu_e$ CC and NC $\pi^0$ events are not constant with the off-axis angle. 
Figure~\ref{fig:ratios} (middle) shows that the number of NC $\pi^0$ events drops more rapidly with the off-axis angle compared to the number of $\nu_e$ CC events.
This occurs for two main reasons: $(i)$ NC $\pi^0$ events predominantly originate from $\nu_{\mu}$ interactions, which are more concentrated near the center of the beam line (see inset in \cref{fig:prism_events:numu}), and $(ii)$ NC $\pi^0$ production requires a higher-energy neutrino to be produced, but the high-energy tail of the neutrino flux reduces quickly with off-axis angle (see \cref{fig:prism_flux}), reducing the likelihood of $\pi^0$ production at large angles.
The behavior shown in \cref{fig:ratios} (middle) provides guidance as to how SBND-PRISM can be used to reduce this particular background.
An electron-neutrino analysis with a significant NC $\pi^0$ contamination could define its key signal region to be in the off-axis regions, where purity increases by 38\%.
In addition, an analysis impacted by NC $\pi^0$ background would benefit from performing a fit in bins of off-axis angle, taking advantage of the NC $\pi^0$ rate reduction in the outer detector region.

A third area of opportunity is in the use of SBND-PRISM to minimize the impact of cross-section modeling uncertainties in analyses investigating models that predict, for example, an off-axis–dependent excess of electron-neutrino interactions.
As shown in \cref{fig:prism_events}, a key feature of the SBND-PRISM effect lies in the correlation between neutrino energy and position in the detector.
The modeling of neutrino–nucleus interactions has large uncertainties, which affect both the estimation of the number of neutrino events and the determination of the neutrino energy spectrum.
Combined with flux uncertainties, these are the main limiting factors in many analyses.
However, neutrino–nucleus interaction uncertainties are independent of the detector geometry, and the on-axis to off-axis flux uncertainties are strongly correlated as they originate in meson decay kinematics.
Therefore, the SBND-PRISM technique offers a promising venue to mitigate the limitations arising from these dominant uncertainties.

An additional handle within this framework comes from the difference between the muon- and electron-neutrino spectra.
The distributions previously shown in \cref{fig:prism_events} are better highlighted in \cref{fig:ratios} (bottom), which shows the ratio of electron- to muon-neutrino CC events in SBND as a function of the off-axis angle.
This ratio has a clear dependence on the off-axis angle, with the number of $\nu_{\mu}$ events dropping more rapidly than the number of $\nu_e$ events.
The ratio changes by about 27\% from the on-axis region to the most off-axis region.
This asymmetry between $\nu_e$ and $\nu_\mu$ provides an extra handle for analyses in which the signal is dominated by the $\nu_{\mu}$ flux while the background is contributed primarily by the $\nu_e$ flux.
In the next section we illustrate how the SBND-PRISM technique can impact systematic uncertainties in a sterile neutrino oscillation search.

\subsection{
An application: Impact of SBND-PRISM on systematic uncertainties}
\label{sec:steriles}

\begin{figure*}
    \centering
    \begin{minipage}[t]{0.358\textwidth}
        \centering
        \includegraphics[width=0.99\textwidth]{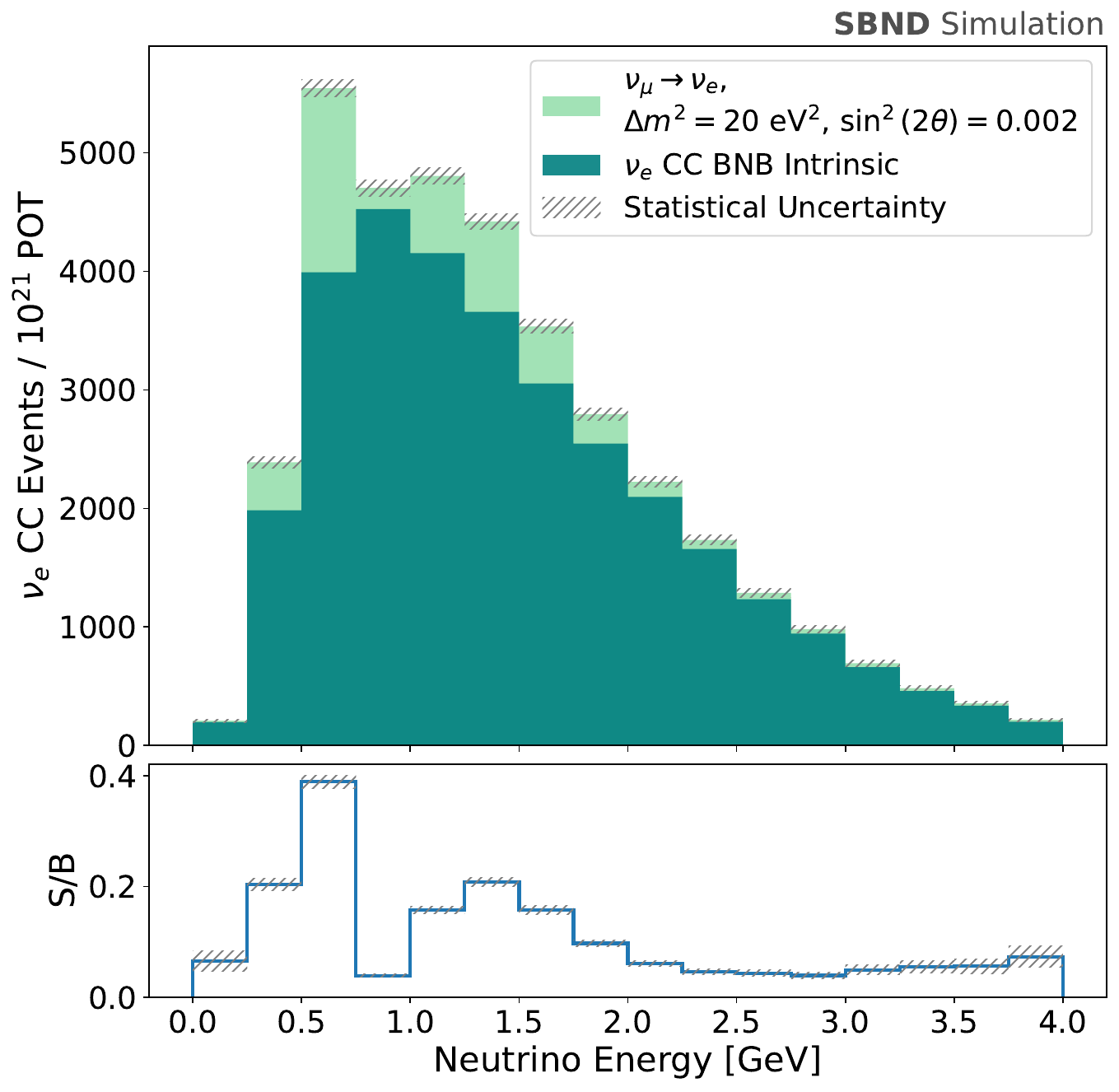}
        \caption{\label{fig:benchmark}
        Expected background events in SBND from electron neutrinos intrinsic to the beam (dark green), and expected signal for $\Delta m^2 = 20$ eV$^2$ and $\sin^2(2\theta) = 0.002$ (light green). Histograms are stacked. Events are binned in true neutrino energy. The blue line shows the ratio between signal and background events in the case of the full SBND volume.}
    \end{minipage}\hfill
    \begin{minipage}[t]{0.61\textwidth}
        \centering
        \includegraphics[width=0.99\textwidth]{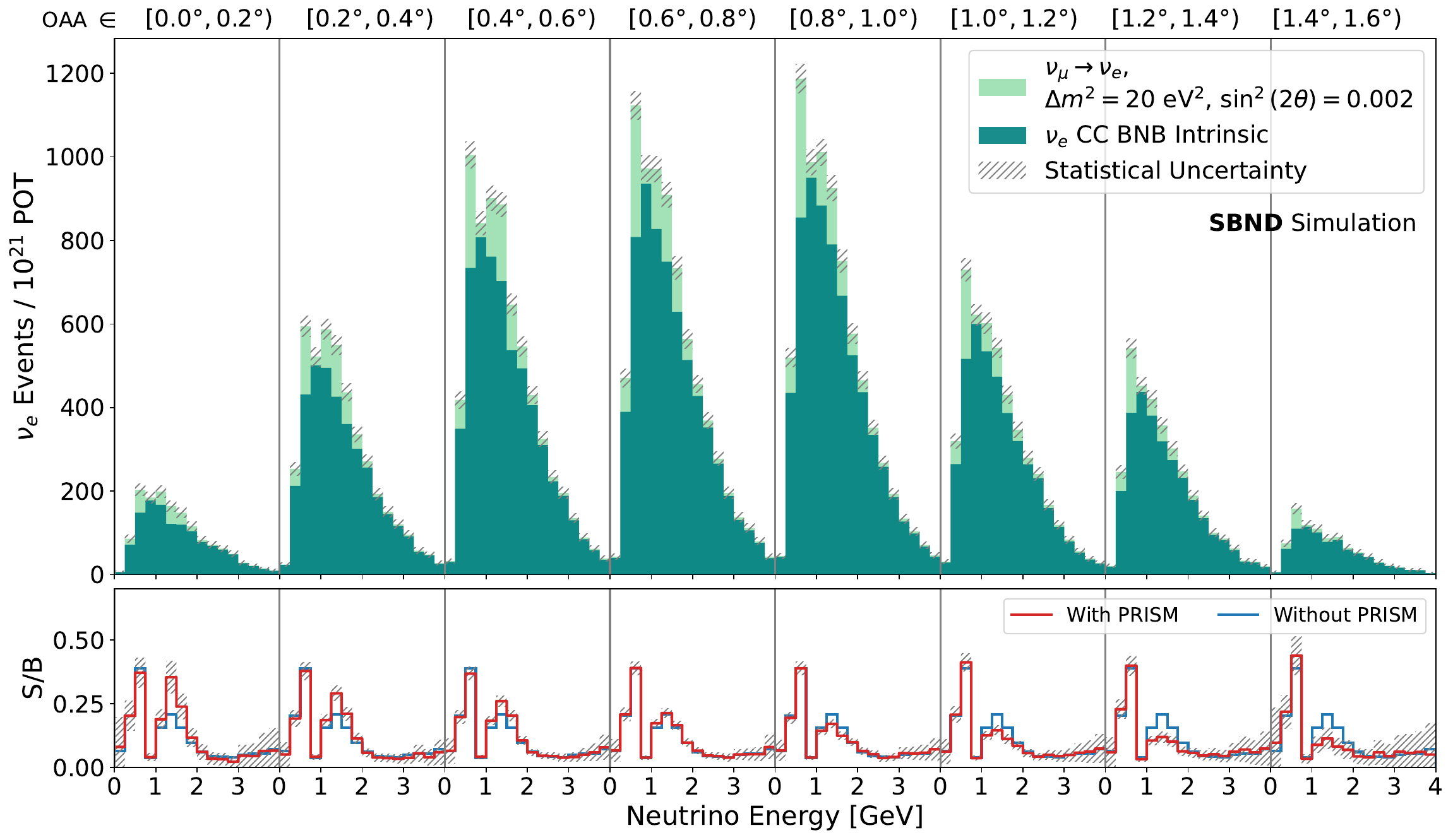}
        \caption{\label{fig:benchmark-prism}Expected background events in SBND-PRISM from electron neutrinos intrinsic to the beam (dark green), and expected signal for $\Delta m^2 = 20$ eV$^2$ and $\sin^2(2\theta) = 0.002$ (light green). Histograms are stacked. Events are binned in true neutrino energy. The red line shows the ratio between the two green distributions, while the blue line shows the ratio in the case of the full SBND volume (same line as in \cref{fig:benchmark}).}
    \end{minipage}
\end{figure*}

In this subsection, we illustrate the SBND-PRISM concept more quantitatively using a specific simplified model of $\nu_e$ appearance from short-baseline oscillations. 
This study is based on generator-level information, without detector simulation or reconstruction effects, and is performed using the SBND spectra alone. It serves to demonstrate the effectiveness of SBND-PRISM in a particular application, motivating future work to fully exploit the potential and assess the impact of this approach.

Since, in the absence of BSM physics, the $\nu_\mu$ fluxes depend on the off-axis angle more dramatically than the $\nu_e$ ones, this model serves as an example of an off-axis-dependent $\nu_e$ excess that can benefit from an SBND-PRISM analysis.

For illustration purposes, we adopt a theoretically simplified scenario, where we assume negligible oscillation effects except for $\nu_\mu\to\nu_e$ appearance given by
\begin{equation}
    P(\nu_\mu\to\nu_e) = \sin^2(2\theta)\sin^2\left(\frac{\Delta m^2 L}{4E}\right),
\end{equation}
where $\theta$ and $\Delta m^2$ are the mixing angle and mass squared difference, respectively, $E$ is the neutrino energy and $L$ is the baseline.

We consider two analysis scenarios: one in which events are binned by off-axis angle (``with SBND-PRISM''), and another in which no angular binning is applied (``without SBND-PRISM''). The procedures used for each scenario are described below.

We only consider the dominant background composed of $\nu_e$ that are intrinsic to the BNB beam. These intrinsic $\nu_e$ events in SBND are shown in the upper panel of \cref{fig:benchmark}.
For comparison, the signal expected from our benchmark scenario of $\nu_\mu \to \nu_e$ oscillation with $\sin^2(2\theta)=0.002$ and $\Delta m^2=20$~eV$^2$ is also shown. These parameters were chosen because higher values of $\Delta m^2$ result in oscillations occurring over shorter baselines, which is where a single near detector like SBND is most sensitive.  This benchmark leads to observable oscillations throughout a large interval of energy in SBND.

\cref{fig:benchmark} shows the relative size of the excess with respect to the background in each energy bin. The shape is driven by the value of $\Delta m^2$ used in the injected oscillation signal.
Although there is a visible excess over the background on the order of 10\% or more, this level of difference is comparable to the leading systematics from flux and cross-section modeling uncertainties~\cite{MicroBooNE:2025nll} which could therefore wash out a signal of this size.
In the worst-case scenario, mismodeling of neutrino-nucleus interaction physics can induce a fake signal~\cite{Coyle:2022bwa}.

The relevance of exploiting SBND-PRISM can be seen in \cref{fig:benchmark-prism}, where we show the same quantities, background and signal for the same benchmark, but now divided in each of the SBND-PRISM angular regions.
In the lower panel, we present the excess with respect to the background for each region, as well as the average excess in blue.
Since cross-section uncertainties at a certain energy would be independent of the geometry of the detector, their impact would affect all regions in a similar way.
Taking the average excess as an example, we can clearly see that it differs from the excess in each region, particularly for the innermost and outermost angular regions.
This is why the use of SBND-PRISM is expected to enhance the experimental sensitivity to BSM scenarios that  exhibit behavior dependent on the off-axis angle, such as the sterile neutrino model.

To further illustrate this capability, we show the effect of SBND-PRISM on a simplified sensitivity calculation for SBND in the sterile neutrino parameter space with, and without, applying SBND-PRISM, under several assumptions about the neutrino-nucleus interaction cross-section uncertainties.

Quantifying uncertainties is a critical step in assessing the effect of SBND-PRISM. We focus on two main sources: flux and cross-section uncertainties.
To assess the uncertainties on the neutrino flux prediction, the flux simulation and systematic uncertainty modeling from the MiniBooNE Collaboration ~\cite{miniboone_flux}, updated to the SBND location, are used.
Flux uncertainties are encoded in a covariance matrix, $E$.
We use a \emph{multisim} technique~\cite{roe}, which consists of generating many MC replicas where each parameter in the flux model is randomly varied from its expected distribution, usually assumed to be a normal distribution with the parameter’s uncertainty as a width.
Simultaneously varying all model parameters allows the correct treatment of correlations between them. $M$ such replicas are created and combined to construct the covariance matrix:
\begin{equation}
\label{eq:multisim}
F_{ij} = \frac{1}{M} \sum_{m = 1}^{M} (n_i^m - n_i^\text{cv})(n_j^m - n_j^\text{cv}),
\end{equation}
where $n_{i}^m$ is the number of events in energy bin $i$ in replica $m$ and $n^\text{cv}_i$ is the expected number of events in bin $i$. 
This approach accounts for correlations in the predicted flux across both energy bins and off-axis angle regions. In particular, correlations between different off-axis regions are significant, with typical values exceeding 85\%, reflecting the common origin of flux uncertainties across the detector.

Addressing neutrino-nucleus interaction uncertainties is a complex challenge, and current modeling falls short of accurately capturing the nuances necessary for interpreting high-statistics neutrino data, as evidenced by both neutrino-nucleus and electron-nucleus scattering data~\cite{Meyer:2016oeg, MINERvA:2019rhx, Ankowski:2019mfd, Ankowski:2020qbe, Ankowski:2020qbe, MINERvA:2020anu, MINERvA:2020zzv, Borah:2020gte, T2K:2020jav, T2K:2021naz, Tomalak:2021qrg, MINERvA:2021owq, NOvA:2021eqi, CLAS:2021neh, Isaacson:2022cwh, MINERvA:2022mnw, Mosel:2023zek, MicroBooNE:2024zwf}. 
For this work, we adopt a model-agnostic approach toward modeling cross-section uncertainties and assume these are uncorrelated across different energy bins and range from 2\% to 200\%, uniformly applied per bin. For comparison, the cross-section uncertainty predicted by GENIE \texttt{v3\_04\_02 AR23\_20i\_00\_000} is around 20\% in almost all energy bins and the average correlation among bins is 40\%. Although this model does not assume a precise knowledge of the underlying physics of neutrino-nucleus interactions, it allows us to explore the impact of cross-section uncertainties on our measurements in a controlled manner.

\begin{figure*}
    \subfloat[]{%
      \includegraphics[width=\columnwidth]{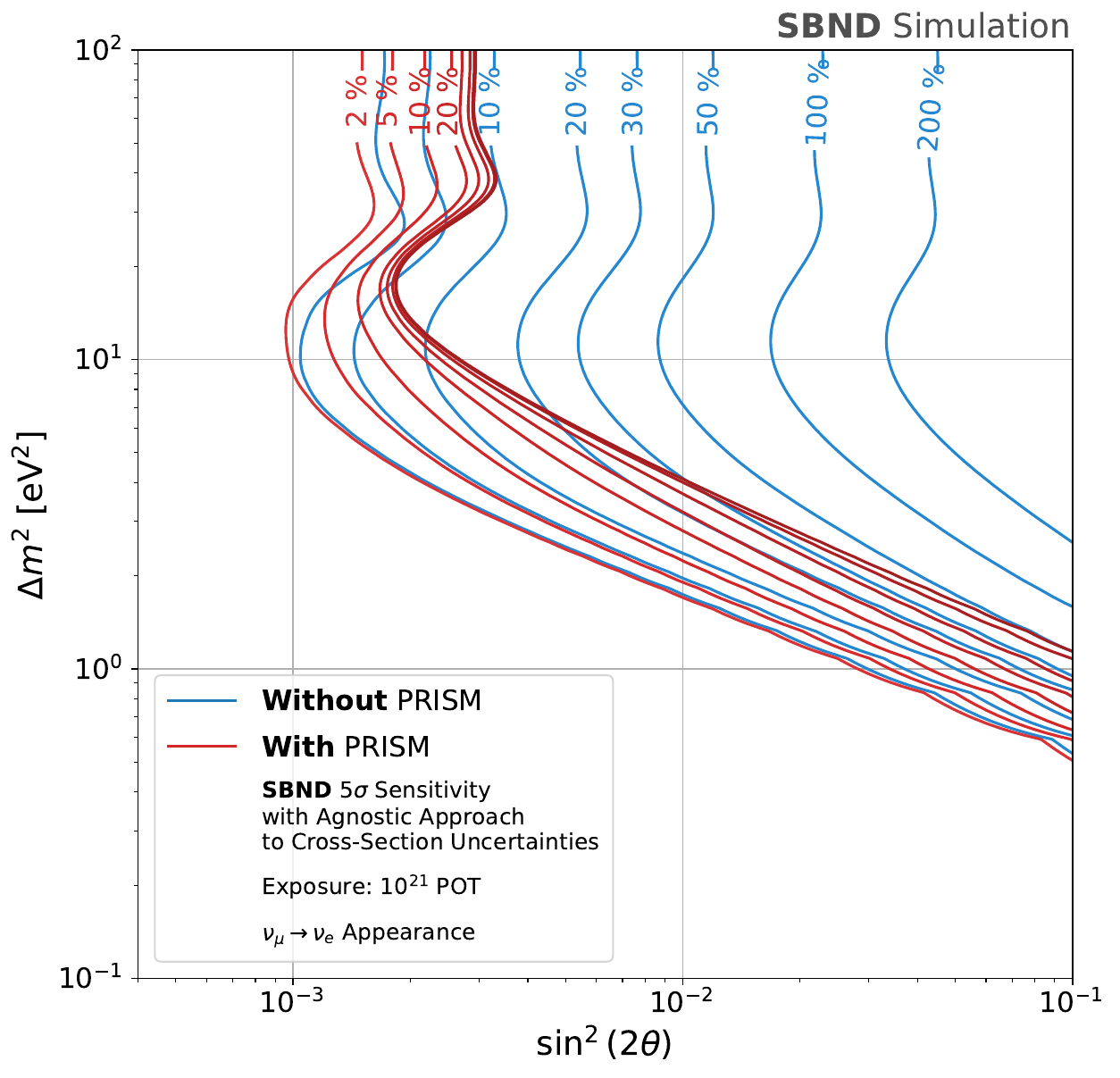}
      \label{fig:sensitivity:all}
    }
    \subfloat[]{%
      \includegraphics[width=\columnwidth]{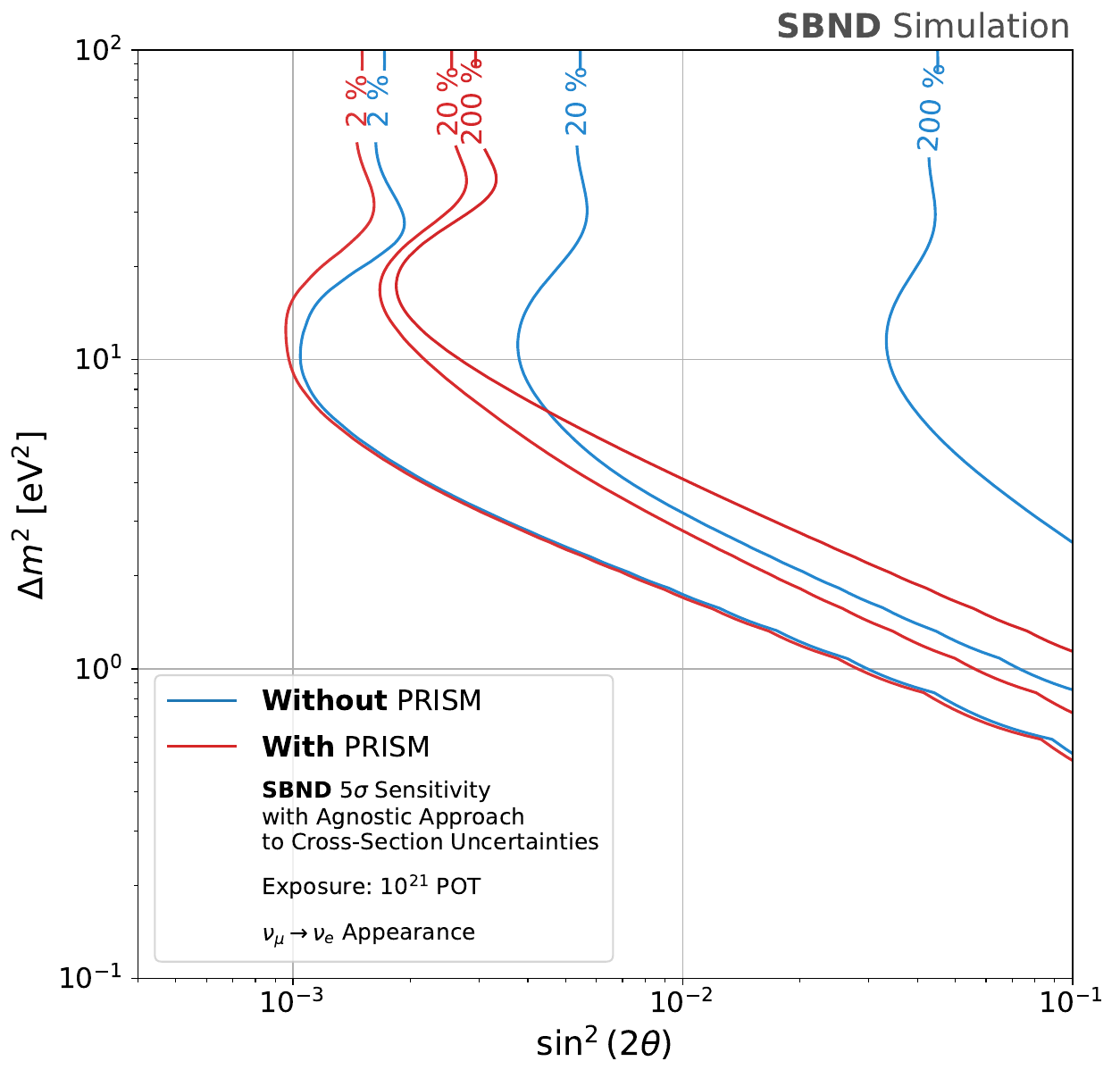}
       \label{fig:sensitivity:some}
    }
    \caption{{\label{fig:sensitivity}
        Sensitivities to $\nu_\mu\to\nu_e$ appearance signals in a simplified scenario in the plane $\Delta m^2$--$\sin^2(2\theta)$, comparing all the different cross-section uncertainties in \protect\subref{fig:sensitivity:all} and only a subset of them for clarity \protect\subref{fig:sensitivity:some}. Different lines represent different cross-section uncertainties, ranging from 2\% to 200\% uncorrelated bin by bin in energy. Only background from intrinsic $\nu_e$ contamination is included.
        }}
\end{figure*}

Systematics are included by combining the BNB flux covariance matrix in \cref{eq:multisim} with cross-section uncertainties to build a full covariance matrix, $C$. 
The sensitivity is calculated by computing a $\chi^2$ surface in the $\Delta m^2$--$\sin^2(2\theta)$ oscillation parameter plane,
\begin{equation}
    \chi^2 = \sum_{i,j}^{\rm bins} (D_i - T_i)(C^{-1})_{ij}(D_j - T_j),
\end{equation}
where $D_i$ is the expected event distribution in the absence of oscillations and $T_i = T_i(\Delta m^2, \sin^2(2\theta))$  is the event prediction for an oscillation signal with mass splitting $\Delta m^2$ and amplitude $\sin^2(2\theta)$.
The labels $i$ and $j$ indicate bins of reconstructed neutrino energy. 
While a complete event reconstruction and selection is beyond the scope of this study, to account for the detector's energy resolution, we smear the electron-neutrino spectrum by $20\%/\sqrt{E/{\rm GeV}}$. MicroBooNE has reported an energy resolution of approximately 10\%--15\%~\cite{PhysRevD.105.112004} for electron neutrinos; however, to be conservative, we adopt a larger smearing. Selection efficiency is not included, but it is not expected to significantly impact the results due to the large event statistics available in each off-axis region.
We evaluate two scenarios: $(i)$ ``without SBND-PRISM'' where the $D^i$ and $T^i$ vectors have 16 entries representing the bin contents of the histograms in \cref{fig:benchmark}, and $(ii)$ ``with SBND-PRISM'' where they have $16 \times 8 = 128$ entries from the histograms in \cref{fig:benchmark-prism}. In the second case, the covariance matrix $C$ contains the flux correlations between energy bins belonging to different off-axis regions. We assume an exposure of $10^{21}$~POT, corresponding to 3 years of SBND data taking. 

In \cref{fig:sensitivity} we present the experimental sensitivities to $\nu_\mu\to\nu_e$ oscillation signals in the plane $\Delta m^2$--$\sin^2(2\theta)$, evaluated with, and without, the SBND-PRISM technique.
The oscillation sensitivities are evaluated by scanning a grid in $\Delta m^2$ and $\sin^2(2\theta)$ and identifying the contour corresponding to a 5$\sigma$ confidence level for two degrees of freedom, under the assumption of Wilks' theorem~\cite{cowan}.
  
We show several cases for the cross-section uncertainties, ranging from 2\% to 200\% uncorrelated uncertainty.
As expected, enlarging the uncorrelated cross-section uncertainties leads to a loss of sensitivity:  
freedom in changing the cross-section energy dependence can accommodate any spectral feature arising from sterile neutrinos.
This effect is significantly mitigated when leveraging SBND-PRISM.
For cross-section systematic uncertainties greater than (15--20)\%, the improvement in the sensitivity due to incorporating SBND-PRISM is substantial. For smaller cross-section uncertainties, the effect is more modest.
As discussed in \cref{sec:fluxes}, the spectral distortions induced by sterile neutrinos have a geometry dependence (\cref{fig:benchmark-prism}). 
This can be traced back to the fact that while the $\nu_\mu\to\nu_e$ signal comes mostly from neutrinos from pion two-body decays, the $\nu_e$ background is dominated by three-body decays of muons and kaons, and has a different neutrino angular distribution.
This is the cause of the SBND-PRISM approach being robust against large cross-section uncertainties.

Furthermore, the use of SBND-PRISM introduces improvements to the experimental sensitivity. Comparing with and without SBND-PRISM for a small 2\% uncertainty in \cref{fig:sensitivity:all}, we can observe a gain in sensitivity of 10\%-20\%, depending on the value of $\sin^2(2\theta)$ when using SBND-PRISM. 

A comprehensive evaluation of the impact of the SBND-PRISM approach on sterile neutrino searches requires a full SBN analysis. While this study is limited to SBND, the techniques and results presented here motivate further exploration within the broader SBN context.

\section{Conclusions}
This paper explores how the angular dispersion of the booster neutrino beam in SBND can enhance its experimental physics capabilities. We present the muon- and electron-neutrino fluxes at the SBND detector as a function of off-axis angle and provide them, along with their associated covariance matrices, for public use~\cite{suppl}. We highlight various opportunities that SBND-PRISM can offer and emphasize how it can be leveraged to perform analyses that are largely insensitive to uncertainties on cross-section modeling. We illustrate the technique's potential through a simplified analysis of $\nu_\mu\to\nu_e$ appearance due to sterile neutrinos in SBND, chosen because the signal and background exhibit distinct dependencies on the off-axis angle. This characteristic makes it particularly well suited to demonstrate the capabilities of SBND-PRISM. We expect that SBND-PRISM will also significantly benefit other BSM searches and neutrino-nucleus interaction studies, broadening the overall physics reach of the experiment.

\begin{acknowledgments}
This document was prepared by the SBND Collaboration using the resources of the Fermi National Accelerator Laboratory (Fermilab), a U.S. Department of Energy, Office of Science, Office of High Energy Physics HEP User Facility. Fermilab is managed by FermiForward Discovery Group, LLC, acting under Contract No. 89243024CSC000002.

The SBND Collaboration acknowledges the generous support of the following organization: 
the U.S. Department of Energy, Office of Science, Office of High Energy Physics; 
the U.S. National Science Foundation;
the Science and Technology Facilities Council (STFC), part of United Kingdom Research and Innovation (UKRI), the UKRI Future Leaders Fellowship, and The Royal Society;
the Swiss National Science Foundation;
the Spanish Ministerio de Ciencia, Innovación y Universidades (MICIU/ AEI/ 10.13039/ 501100011033) under Grants No. PRE2019-090468, No. CNS2022-136022, No. RYC2022-036471-I, and No. PID2023-147949NB-C51 and C53 and Comunidad de Madrid (PEJ-2023-AI/COM-28399);
the European Union’s Horizon 2020 research and innovation program under GA No. 101004761 and the Marie Sk\l{}odowska-Curie Grant Agreements No. 822185, No. 101081478, and No. 101003460;
the São Paulo Research Foundation 1098 (FAPESP), the National Council of Scientific and Technological Development (CNPq), and Ministry of  Science, Technology and Innovations-MCTI of Brazil. 
\end{acknowledgments}

\section*{Data Availability}
The data that support the findings of this article are openly available~\cite{suppl}.


\bibliography{apssamp}

\end{document}